\documentclass[longauth,structabstract]{aa} 
\usepackage{graphicx}
\usepackage{txfonts}
\usepackage{natbib}
\usepackage{xspace}
\def\deg{$^\circ$\xspace}
\def\mic{$\mu$m\xspace}
\def\Msun{M$_\odot$\xspace} 

\begin{document}
   \title{{\it Herschel} images of Fomalhaut\thanks{{\it Herschel} is an ESA space observatory with science instruments provided by European-led Principal Investigator consortia and with important participation from NASA.}}

   \subtitle{An extrasolar Kuiper Belt at the height of its dynamical activity}

   \author{B. Acke\inst{1}\fnmsep\thanks{Postdoctoral Fellow of the Fund for Scientific Research, Flanders}
   \and M. Min\inst{2,3}
   \and C. Dominik\inst{3,4}
   \and B. Vandenbussche\inst{1}
   \and B. Sibthorpe\inst{5}
   \and C. Waelkens\inst{1}
   \and G. Olofsson\inst{6}
   \and P. Degroote\inst{1} 
   \and K. Smolders\inst{1}\fnmsep\thanks{Aspirant Fellow of the Fund for Scientific Research, Flanders}
   \and E. Pantin\inst{7} 
   \and M. J. Barlow\inst{8}
   \and J. A. D. L. Blommaert\inst{1} 
   \and A. Brandeker\inst{6}
   \and W. De Meester\inst{1} 
   \and W. R. F. Dent\inst{9}
   \and K. Exter\inst{1} 
   \and J. Di Francesco\inst{10}
   \and M. Fridlund\inst{11}
   \and W. K. Gear\inst{12} 
   \and A. M. Glauser\inst{5,13} 
   \and J. S. Greaves\inst{14} 
   \and P. M. Harvey\inst{15} 
   \and Th. Henning\inst{16}
   \and M. R. Hogerheijde\inst{17} 
   \and W. S. Holland\inst{5} 
   \and R. Huygen\inst{1} 
   \and R. J. Ivison\inst{5,18} 
   \and C. Jean\inst{1} 
   \and R. Liseau\inst{19} 
   \and D. A. Naylor\inst{20} 
   \and G. L. Pilbratt\inst{11}
   \and E. T. Polehampton\inst{20,21} 
   \and S. Regibo\inst{1} 
   \and P. Royer\inst{1} 
   \and A. Sicilia-Aguilar\inst{16,22}
   \and B.M. Swinyard\inst{21}}

   \institute{Instituut voor Sterrenkunde, KU Leuven, Celestijnenlaan 200D, B-3001 Leuven, Belgium\\
   \email{bram@ster.kuleuven.be}
   \and
   Astronomical Institute, Utrecht University, Princetonplein 5, 3584 CC Utrecht, the Netherlands.
   \and
   Astronomical Institute Anton Pannekoek, University of Amsterdam, Kruislaan 403, 1098 SJ Amsterdam, the Netherlands.
   \and
   Afdeling Sterrenkunde, Radboud Universiteit Nijmegen, Postbus 9010, 6500 GL Nijmegen, the Netherlands.
   \and
   UK Astronomy Technology Centre, Royal Observatory Edinburgh, Blackford Hill, EH9 3HJ, UK.
   \and
   Department of Astronomy, Stockholm University, AlbaNova University Center, 106 91 Stockholm, Sweden.
   \and
   Laboratoire AIM, CEA/DSM-CNRS-Universit{\'e} Paris Diderot, IRFU/Service d'Astrophysique, Bat. 709, CEA-Saclay, 91191 Gif-sur-Yvette Cedex, France.
   \and
   Department of Physics and Astronomy, University College London, Gower St, London WC1E 6BT, UK.
   \and
   ALMA, Alonso de C{\'o}rdova 3107, Vitacura, Santiago, Chile.
   \and
   National Research Council of Canada, Herzberg Institute of Astrophysics, 5071 West Saanich Road, Victoria, BC, V9E 2E7, Canada.
   \and
   ESA Research and Science Support Department, ESTEC/SRE-S, Keplerlaan 1, NL-2201AZ, Noordwijk, the Netherlands.
   \and
   School of Physics and Astronomy, Cardiff University, Queens Buildings The Parade, Cardiff CF24 3AA, UK.
   \and
   Institute of Astronomy, ETH Zurich, 8093 Zurich, Switzerland.
   \and
   School of Physics and Astronomy, University of St Andrews, North Haugh, St Andrews, KY16 9SS, UK.
   \and
   Department of Astronomy, University of Texas, 1 University Station C1400, Austin, TX 78712, USA.
   \and
   Max-Planck-Institut f{\"u}r Astronomie, K{\"o}nigstuhl 17, D-69117 Heidelberg, Germany.
   \and
   Leiden Observatory, Leiden University, PO Box 9513, 2300 RA, Leiden, the Netherlands.
   \and
   Institute for Astronomy, University of Edinburgh, Blackford Hill, Edinburgh EH9 3HJ, UK.
   \and
   Earth and Space Sciences, Chalmers University of Technology, SE-412 96 Gothenburg, Sweden.
   \and
   Institute for Space Imaging Science, University of Lethbridge, Lethbridge, Alberta, T1J 1B1, Canada.
   \and
   Space Science and Technology Department, Rutherford Appleton Laboratory, Oxfordshire, OX11 0QX, UK.
   \and
   Departamento de F{\'i}sica Te{\'o}rica, Universidad Aut{\'o}noma de Madrid, Cantoblanco 28049, Spain.
   }

   \date{Received ; accepted}

 
  \abstract
   {Fomalhaut is a young ($2 \pm 1 \times 10^8$ years), nearby (7.7~pc), 2~\Msun star that is suspected to harbor an infant planetary system, interspersed with one or more belts of dusty debris. }
   {We present far-infrared images obtained with the \textit{Herschel} Space Observatory with an angular resolution between 5.7\arcsec\ and 36.7\arcsec\ at wavelengths between 70~\mic and 500~\mic. The images show the main debris belt in great detail. Even at high spatial resolution, the belt appears smooth. The region in between the belt and the central star is not devoid of material; thermal emission is observed here as well. Also at the location of the star, excess emission is detected. We aim to construct a consistent image of the Fomalhaut system.}
   {We use a dynamical model together with radiative-transfer tools to derive the parameters of the debris disk. We include detailed models of the interaction of the dust grains with radiation, for both the radiation pressure and the temperature determination. Comparing these models to the spatially resolved temperature information contained in the images allows us to place strong constraints on the presence of grains that will be blown out of the system by radiation pressure. We use this to derive the dynamical parameters of the system.}
   {The appearance of the belt points towards a remarkably active system in which dust
grains are produced at a very high rate by a collisional cascade in a narrow region
filled with dynamically excited planetesimals. Dust particles with sizes below the blow-out size are abundantly present. The equivalent of 2000 one-km-sized comets are destroyed every day, out of a cometary reservoir amounting to 110 Earth masses. From comparison of their scattering and thermal properties, we find evidence that the dust grains are fluffy aggregates, which indicates a cometary origin. The excess emission at the location of the star may be produced by hot dust with a range of temperatures, but may also be due to gaseous free-free emission from a stellar wind.}
   {}


\maketitle
%

\section{Introduction}

\begin{figure}
\centering
\includegraphics[width=\columnwidth]{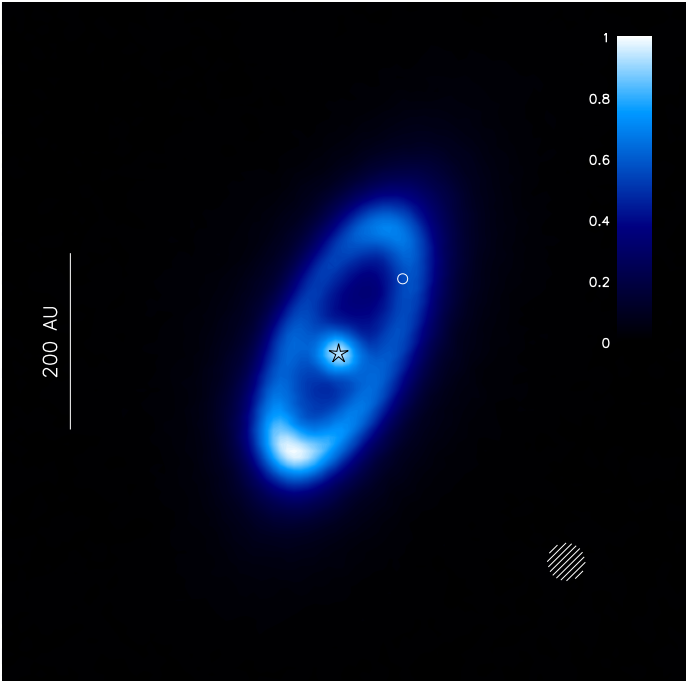}
\caption{{\it Herschel} PACS 70~\mic image of Fomalhaut. The positions of the star and of the planet candidate Fomalhaut~b are indicated. The hatched circle indicates the spatial resolution (5.7\arcsec). The black-blue-white color scale is linear. North is up, east to the left.}
           \label{fom_image_70}%
\end{figure}

\object{Fomalhaut} was one of the first main-sequence stars shown to have an infrared excess
caused by dust grains in orbit around the star \citep{aumann85}. Further studies of this phenomenon have shown that approximately 15\% of solar-like main-sequence stars have a detectable amount of debris \citep{bryden06}. Belts of debris typically reside in planetary systems at locations where no planets have formed because either the formation time scale for planets was too long, such as in the Solar System's Kuiper Belt, or because other planets in the system have stirred up the belt before a planet could be formed, such as in the Solar System's Asteroid Belt \citep{wyatt08,krivov10}. 

Fomalhaut's disk was previously imaged from the ground at 450~\mic and 850~\mic, and from space with the {\it Spitzer} and the {\it Hubble} Space Telescopes \citep{holland03, stapelfeldt04,kalas05}. A planet candidate has been reported in the disk, though 3$^\mathrm{rd}$ epoch confirmation of this object is still uncertain \citep{kalas08,kalas11,janson12}.
The most prominent feature in the disk is the main belt, which is located outside of 130
AU. The center of this belt is shifted with respect to the star, implying significant
eccentricity, probably caused by gravitational interaction with one or more planets \citep{kalas05,kalas08,chiang09}.

\section{Observations}

\subsection{{\it Herschel} images}

Far-infrared images of Fomalhaut have been obtained with the PACS \citep{poglitsch10} and
SPIRE \citep{griffin10,swinyard10} instruments aboard the {\it Herschel} Space Observatory \citep{pilbratt10} in five photometric filters centered around 70~\mic, 160~\mic (PACS), 250~\mic, 350~\mic and 500~\mic (SPIRE). The Observation IDs are 1342211953, 1342211954 (PACS) and 1342195939 (SPIRE). The full width at half maximum of the point spread functions (PSFs) indicate a spatial resolution of the observations of 5.7\arcsec, 11.3\arcsec, 18.1\arcsec, 25.1\arcsec\ and 36.7\arcsec\ for the five bands respectively. This corresponds to 43, 86, 140, 190 and 280~AU at the distance to Fomalhaut. 

The PACS data were processed in the {\it Herschel} Interactive Processing Environment \citep[HIPE,][]{ott10} version 6.0, applying the standard pipeline steps. The flux conversion was done using version 5 of the response calibration. Signal glitches due to cosmic-ray impacts were masked using the PACS {\em photMMTDeglitching} task in HIPE on the detector timeline. Then a first coarse map was projected. A high-pass filter was applied to remove the low frequency drifts. To avoid the introduction of artefacts, we used the coarse map to mask out the disk from the detector timeline prior to the application of the filter. The scale of the high-pass filter was 15 frames for the blue maps (70~\mic), and 25 frames for the red maps (160~\mic). The detector time-series signals were then summed up into a map using the PACS {\em photProject} task. The pixel scale for the 70~\mic map was set to 1\arcsec, while the scale for the 160~\mic map was 2\arcsec. The source was covered in two independent observations, scanning the sky in two orthogonal directions.  We combined the two detector time series and projected these together into the final maps.

Also the SPIRE data were reduced using HIPE. The high signal-to-noise ratio of these data allowed us to created over-sampled maps with pixel scales of 3\arcsec, 4.2\arcsec\ and 7\arcsec\ in the 250~\mic, 350~\mic and 500~\mic bands respectively, equating to approximately 6 samples per beam FWHM.  The standard pipeline script provided with HIPE was used to reduce the data, with the {\em naiveMapper} task being used to construct the maps. A simple linear baseline subtraction was the only filtering performed on these data.

Table~\ref{image_par} summarizes the pixel scale, total map integration time and signal-to-noise ratio of the projected {\it Herschel} images. We also indicate the integration time per pixel, which is the total time a pixel in the central part of the projected map was seen by a detector pixel.

\begin{figure}
\centering
\includegraphics[width=\columnwidth]{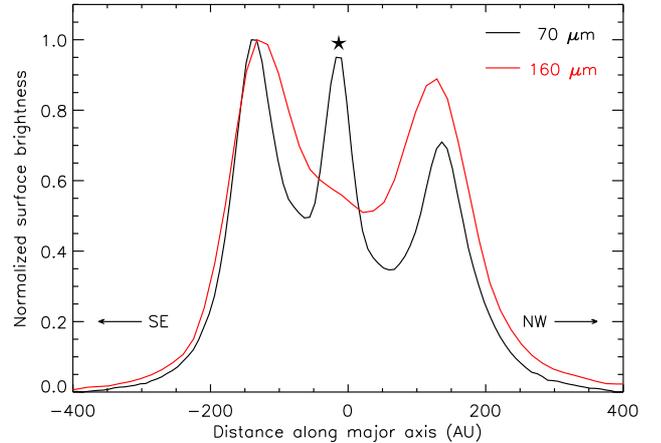}
\caption{Cut through the {\it Herschel}/PACS images along the belt's major axis on the sky. The surface brightness profiles are normalized by the maximum values indicated in Table~\ref{image_par}. The off-center position of the star is indicated. The belt's peak positions are located closer to the belt center at 160\,\mic, which is a consequence of the reduced angular resolution (see Sect.~\ref{bg}).}
           \label{fom_cut}%
\end{figure}

\begin{figure*}
\centering
\includegraphics[height=0.92\textheight]{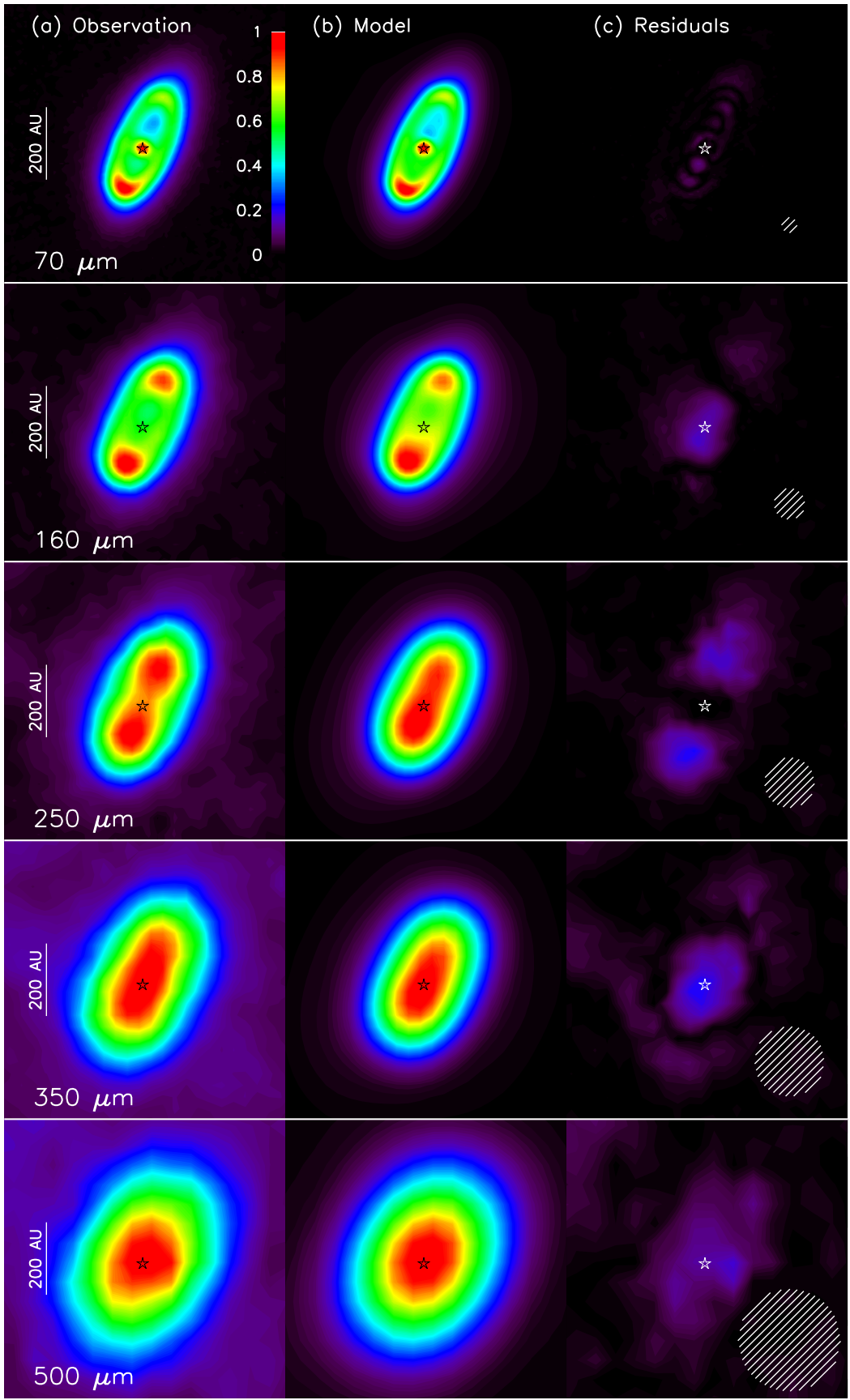}
\caption{{\it Herschel} far-infrared images of Fomalhaut. {\em (a)} Observations, {\em (b)} Model images, {\em (c)} Absolute value of the residuals. The hatched circles indicate the spatial resolution. The color scale is linear rainbow to increase contrast at low surface brightness. The scale ranges from zero to maximum surface brightness (see Table~\ref{image_par}) of the image in columns {\em (a)} and {\em (b)}; In column {\em (c)}, the color scale is the same as in column {\em (a)}. North is up, east to the left.}
           \label{fom_images}%
\end{figure*}

Figs.~\ref{fom_image_70}, \ref{fom_cut} and \ref{fom_images}  show that the system is well resolved at 70~\mic, barely resolved at 500~\mic, and marginally resolved at intermediate wavelengths. A first qualitative look at the images shows that the thermal emission of the dust in the belt around Fomalhaut appears smooth and not clumpy, contrary to what was expected by \citet{wyatt02}. The smoothness hints at a high dust replenishment rate by numerous collisions, rather than sporadic replenishment by rare collisional events. The apparent azimuthal brightness variations are due to geometric projection effects. Moreover, distances from the belt to the star vary along the belt, as it is eccentric. The southern ansa is located closer to the star than the northern ansa, and is therefore warmer and brighter. This is the so-called pericenter glow, first observed in a debris disk by \citet{telesco00}. The planet candidate Fomalhaut~b may produce local density enhancements. However, there is no evidence of an excess or a lack of dust emission near the position of Fomalhaut~b, nor at any other location along the belt.

In the 70~\mic image, the star shows up as an unresolved point source. The flux of this central source, however, exceeds the stellar photospheric flux. This is in agreement with excess flux close to the star previously found at near-infrared wavelengths and attributed to hot dust \citep{absil09}. The region between the unresolved central source and the belt is not devoid of dust either; the 70~\mic image shows emission in this inner disk as well. The dust particles in this region either originated in the belt and spiraled in past the orbit of Fomalhaut~b, or were released from an unresolved inner belt and moved outwards.

\begin{table*}
\caption{{\it Herschel} images.}
\label{image_par}      
\centering                          
\begin{tabular}{llllll}        
\hline\hline                 
Image & Pixel$^\mathrm{a}$ (\arcsec) & \multicolumn{2}{c}{Integration time (s)} & SB$^\mathrm{b}$ & SNR$^\mathrm{b}$ \\    
 & & Total map & Per pixel$^\mathrm{c}$ & (mJy/\arcsec$^2$) &  \\
\hline                        
PACS 70~\mic     & 1   & 10956 & 15.6 & 16.5  & 77 \\
PACS 160~\mic    & 2   & 10956 & 16.7 &  6.28 & 59 \\
SPIRE 250~\mic   & 3   &  2906 &  2.1 &  1.90 & 32 \\
SPIRE 350~\mic   & 4.2 &  2906 &  2.8 &  0.59 & 21 \\
SPIRE 500~\mic   & 7   &  2906 &  3.6 &  0.19 & 27 \\
\hline                                   
\end{tabular}
\begin{list}{}{}
\item[$^{\mathrm{a}}$] Pixel size of the projected image. 
\item[$^{\mathrm{b}}$] Surface brightness and signal-to-noise ratio in the brightest pixel of the image.
\item[$^{\mathrm{c}}$] In the central part of the image where the source is located.
\end{list}
\end{table*}

\begin{table*}
\caption{Geometric properties of the elliptical belt.}
\label{belt_par}      
\centering                          
\begin{tabular}{llll}        
\hline\hline                 
 & Optical$^\mathrm{a}$ & 70~\mic & 160~\mic \\    
\hline                        
Semi-major axis $a$ (AU)$^\mathrm{b}$ & $141 \pm 2$ & $137.5 \pm 0.9$ & $125 \pm 3^\mathrm{c}$ \\
Eccentricity $e$ & $0.11 \pm 0.01$ & $0.125 \pm 0.006$ & $0.17 \pm 0.07$ \\
Inclination $i$ (\deg) & $65.6 \pm 0.4$ & $65.6 \pm 0.5$ & $64 \pm 2$ \\
Longitude of ascending node $\Omega$ (\deg) & $156.0 \pm 0.3$ & $156.9 \pm 0.5$ & $160 \pm 2$ \\
Argument of periastron $\omega$ (\deg) & $31 \pm 6$ & $1 \pm 6$ & $\dots^\mathrm{d}$ \\
Offset between star and ring center (AU)$^\mathrm{e}$ & $15 \pm 1$ & $17.2 \pm 0.9$ & $\dots^\mathrm{d}$ \\
\hline                                   
\end{tabular}
\begin{list}{}{}
\item[$^{\mathrm{a}}$] \citet{kalas05}.
\item[$^{\mathrm{b}}$] At peak surface brightness.
\item[$^{\mathrm{c}}$] Spatial resolution of the 160~\mic image is insufficient to derive a reliable semi-major axis.
\item[$^{\mathrm{d}}$] Star not visible at 160~\mic.
\item[$^{\mathrm{e}}$] In the plane of the belt.
\end{list}
\end{table*}

\subsection{Belt geometry \label{bg}}

The geometric properties of the belt were derived from the PACS 70~\mic and 160~\mic images in the following way. Along each pixel row and column intersecting the belt, the coordinates of the pixel with the highest surface brightness are determined. In this way, the position of the ring in the image is captured by a set of pixel coordinates $(x,y)$ along the ring. An ellipse is fitted to these coordinates, which yields the major axis, the minor axis, and the position angle of the belt as it appears on the sky. The position of the star, relative to the belt center, is measured from the PACS 70~\mic image by fitting a two-dimensional Gaussian curve to the central unresolved point source. From the parameters of the apparent ellipse, the stellar position, and the known distance to Fomalhaut, the orbital elements of the belt are computed: the semi-major axis $a$, the eccentricity $e$, the inclination $i$, the longitude of the ascending node $\Omega$, and the argument of periastron $\omega$.

We repeat this ellipse-fitting procedure 1000 times by adding random noise in each spatial pixel (spaxel) of the image. The magnitude of this added noise is equal to three times the pixel-to-pixel noise of the original image, estimated by taking the standard deviation of the image outside the location of the Fomalhaut system on the sky. As a result of this added noise, the set of belt locations $(x,y)$ are not exactly the same for two different simulations (although several pixels may be in common). Also the measured stellar position varies. Each simulation will therefore yield slightly different parameters. Our final set of belt parameters are the mean values, and the errors the standard deviations, of the outcome of these 1000 simulations.
The same procedure is applied to derive the belt parameters from the PACS 160~\mic image. Because the star is not visible in this image, however, we use the relative stellar position measured in the 70~\mic image.

The geometric properties of the belt are summarized in Table~\ref{belt_par}. They are similar to those derived from optical scattering images \citep{kalas05}, apart from the argument of periastron $\omega$. We find that the star is located almost perfectly on the apparent major axis of the belt on the sky, leading to $\omega = 1^\circ \pm 6^\circ$. \citet{kalas05} do not explicitly mention the value, but $\omega$ can be estimated from the measured difference ($\delta=14$\deg) between their longitude of the nodal line ($\Omega = 156$\deg) and their belt center-star offset direction (170\deg), deprojected for the inclination ($i=65.6$\deg): $\omega = \arctan \left( \tan{\delta}\,/\cos{i} \right) = 31 \pm 6$\deg.  The reason for this discrepancy is unclear to us, but we are confident in our value. Our {\it Herschel} 70~\mic image is unique in that it captures both the stellar light and the thermal dust emission simultaneously, in contrast to the {\it Hubble} Space Telescope (HST) images, where the direct starlight had to be blocked to be able to detect the light scattered by the belt. Our measurement of $\omega$ is therefore more direct. To elaborate on this issue, we have tested both values of $\omega$ with our radiative-transfer model (see Sect.~\ref{modeling}). In the case of $\omega = 31$\deg, the pericenter glow in the model images is clearly {\it off} the major axis of the belt, towards the South, unlike the observed on-axis pericenter glow\footnote{The model images are convolved with the PACS 70~\mic point spread function before comparison with the observations. All asymmetries related to the latter are hence taken into account.}. The on-axis position of both star and pericenter glow, lead us to conclude that $\omega$ must be close to zero degrees.

Although the HST optical images have a spatial resolution much higher than that of {\it Herschel}/PACS at 70~\mic (0.5~AU vs 43~AU at the distance of Fomalhaut), the error bars on the belt parameters are comparable. This is mainly due to the superb signal-to-noise ratio of the PACS images (see Table~\ref{image_par}).

The belt properties derived from the 160~\mic image are less certain. The derived semi-major axis (125~AU) is significantly smaller than the one derived from the 70~\mic image (137.5~AU). We ascribe the difference to the lower angular resolution; The larger PSF at 160~\mic makes the belt appear smaller, because it {\em squeezes} the locations of peak surface brightness inwards.

\subsection{Spectral energy distribution}

\begin{table*}
\caption{{\it Herschel} photometry of Fomalhaut.}
\label{fom_phot}      
\centering                          
\begin{tabular}{llcccc}        
\hline\hline                 
Band & Integrated & \multicolumn{4}{c}{Flux contribution (\%)$^{\mathrm{b}}$}  \\
     & flux (Jy)$^{\mathrm{a}}$  & Point source & Inner disk & Belt & Blow-out grains \\
\hline                        
PACS 70~\mic   & $10.8 \pm 0.9^{\mathrm{c}}$  & 5 & 21 & 74 & 27 \\
PACS 160~\mic  & $6.2 \pm 0.6$   & 2 & 23 & 75 & 7 \\
SPIRE 250~\mic & $2.7 \pm 0.3$   & 2 & 25 & 73 & 3 \\
SPIRE 350~\mic & $1.1 \pm 0.1$   & 2 & 27 & 71 & 2 \\
SPIRE 500~\mic & $0.50 \pm 0.05$ & 2 & 29 & 69 & 2 \\
\hline                                   
\end{tabular}
\begin{list}{}{}
\item[$^{\mathrm{a}}$] Within a circular aperture with a radius of 52\arcsec\ (400~AU) around the geometric center of the belt.
\item[$^{\mathrm{b}}$] From the best-fit model.
\item[$^{\mathrm{c}}$] The excess flux at the location of the star is $50 \pm 10$\% of the stellar flux at 70~\mic, or $0.17 \pm 0.02$ Jy.
\end{list}
\end{table*}

{\it Herschel} photometry was derived from the images described above, by integrating the flux within a circular aperture around the geometric center of the belt on the sky. We used an aperture radius of 52\arcsec, i.e. 400~AU at the distance of Fomalhaut (Table~\ref{fom_phot}). The mentioned uncertainties consist of the statistical errors, amounting to $\sim$10\% of the measured flux. According to the SPIRE and PACS documentation available from the {\it Herschel} Science Center web pages\footnote{http://herschel.esac.esa.int/}, the absolute calibration error on the photometry is another 10\%.

Additional photometry of the Fomalhaut system was extracted from several online catalogues, available from http://vizier.ustrasbg.fr/viz-bin/VizieR. Stellar photosphere models \citep{castelli03} were fitted to the
optical Geneva photometry \citep{rufener88}. We have interpolated between the grid models of \citet{castelli03}, according to the method explained in \citet[][their Sect.~2]{degroote11}. The derived fundamental parameters of the star are an effective temperature T$_\mathrm{eff}$ = $8600\pm200$~K, a surface gravity $\log{g} = 4.1\pm0.2$~dex, and an interstellar reddening E(B$-$V) $< 0.08$. The photometric angular diameter of the star is $2.20\pm0.04$ milli-arcsec, in perfect agreement with interferometric measurements \citep{difolco04}. The infrared photometry is taken from the AKARI/IRC and FIS All-Sky Survey point source catalogues \citep{ishihara10,yamamura10}, the ESO catalogue “IR photometry of calibrator stars” \citep{vanderbliek96}, and the {\it Spitzer} Space Telescope MIPS catalogue \citep{su06}. The two sub-mm measurements, at 450~\mic and 850~\mic, are consistent with the SPIRE photometry \citep{holland03}. The spectral energy distribution (SED) of the Fomalhaut system is shown in Fig.~\ref{fom_sed}.

\section{Modeling \label{modeling}} 

\subsection{Three components}

We use a three-component model to reproduce the detected thermal dust
emission. A central, unresolved component responsible for the stellar plus excess emission at the location of the star, a narrow ring of dust produced in a collisional cascade for the belt, and a power-law surface density to reproduce residual emission inside the belt.

We assume that the dust in the belt originates from a narrow source region, a ring between radii $r_1$ and $r_2$ in which dust grains are produced through collisions with a size distribution according to the equilibrium cascade: $f(s) \propto s^{-3.5}$, with $s$ defined as the radius of the grain \citep{dohnanyi69}. The particles are assumed to be released from a parent body on an orbit inside the ring. For small grains, their orbits are modified by stellar radiation forces on the particles. The modification depends on the ratio $\beta = F_\mathrm{rad}/F_\mathrm{grav}$ of radiation and gravitational forces acting on the particles. Particles with $\beta < 0.5$ will remain in bound orbits around the star, with periastrons $r_\mathrm{p}$ equal to their release points in the ring, and forced eccentricities $e = \beta / (1 - \beta)$ so that the semi-major axes become $a = r_\mathrm{p} / (1 - e)$. Hence, particles with increasing $\beta$ will find themselves in orbits with increasing eccentricities and apastron distances. Particles with a given $\beta$ will produce a surface density distribution inversely proportional to $2 \pi^2 a \sqrt{a^2 e^2 - (r - a)^2}$ in the region $a (1 - e) < r < a (1 + e)$.

Particles with $\beta > 0.5$ will be on hyperbolic orbits, and gradually leave the system. These so-called blow-out particles are constantly replenished by the collisional cascade and contribute at low levels to the surface density outside of the ring. We assume that the small-grain extension of the collisional cascade power-law is replenished every $t_\mathrm{repl}$ years, which provides a scaling factor for the surface density of the blow-out particles. The replenishment time is a free parameter of the model.

For particles in the inner-disk region, we take a power-law size distribution with $f(s) \propto s^{-3}$. Such a distribution will result from drawing grains by means of Poynting-Robertson (PR) drag from the belt \citep{burns79}, as explained in Appendix~\ref{appA}. Moreover, we remove all particles with $\beta > 0.5$ from the inner disk, since these will be blown out quickly. The size distribution in the inner disk is not well constrained by the {\it Herschel} images, however. Besides PR drag, evaporating comets and additional debris belts closer to the star are possible sources of dust grains in the inner disk between star and outer belt that could have very different size distributions. Including only PR drag and assuming that no collisions occur in the PR stream inward of the source ring, a constant surface density in the inner disk is expected (see Appendix~\ref{appA}), resulting in a surface brightness profile that is heavily peaked towards the central star but this is inconsistent with our observations. A planetary system inside the ring could clear out the particles as they drift in, creating a surface density profile that decreases towards the star. We model this distribution using a power-law surface density where the exponent is a free parameter.

The central source in the 70~\mic image is modeled with a point source, convolved with the
PSF of the PACS instrument at this wavelength. The model hence consists of three geometric components: a belt, an inner disk and a central point source.

\subsection{The dust model}

\begin{figure}
\centering
\includegraphics[width=\columnwidth]{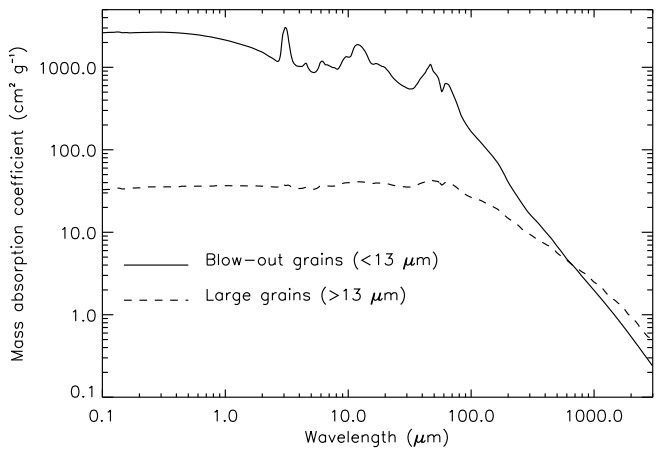}
\caption{The mass absorption coefficients of the dust grain mixture used in the models, averaged for grains with sizes below (full line) and above (dashed line) the blow-out size (13\,\mic). See text for details.}
           \label{opacities.ps}%
\end{figure}

The dust particles in our simulations are mixtures of common astronomical species, including water ice, following the composition proposed by \citet{min11}. This is a mixture consistent with the solar abundances derived by \citet{grevesse98}. The resulting composition contains by mass 32\% silicates, 10\% iron sulfide, 13\% amorphous carbon, and 45\% water ice.

To simulate inhomogeneous grains we mix the various materials using effective medium theory. For this we use the Bruggeman mixing rule \citep{bruggeman35,voshchinnikov07}. We simulate porosity of the grains by adding a fraction of vacuum to the mixture. The grain shape is described by a distribution of hollow spheres \citep[DHS,][]{min05} with an irregularity parameter $f_\mathrm{max} = 0.8$. With a mild porosity of 25\% vacuum, we find that the blow-out size for grains in the Fomalhaut system is 13\,\mic. We tested various values of the porosity and also dust particles without ice to mimic the effect of photosputtering \citep{grigorieva07}. 

For the initial size distribution formed by the collisional cascade we use a power law, as stated above, a minimum grain size of 1\,\mic, and a nominal maximum grain size of 5000\,\mic. The latter value is not well constrained by the observations. The size distribution at each location in the disk is computed following the dynamical model described above. The averaged mass absorption coefficients for grains formed in the collisional cascade with sizes below 13\,\mic, i.e. the blow-out grains, and the averaged opacities for grains larger than the blow-out size are plotted in Fig.~\ref{opacities.ps}. Note that these opacities do not necessarily represent the averaged opacities at any location in the disk since the size distribution is significantly altered by the radiation dynamics.

\subsection{The model grid}

We set up a grid of models with varying inner and outer radius for the ring where the planetesimals collide. We fixed the value of $r_1$ equal to the inner-edge radius of 133~AU that was well determined based on the HST images \citep{kalas05}. We also tested a slightly different $r_1$ of 123 and 143~AU. For the outer radius of the ring, we took $r_2 = r_1 + [2,5,10,20,30,40]$~AU. For the inner disk, we tested surface-density power laws proportional to $r^0$ (PR), $r^1$ and $r^2$. The outer edge of the inner disk is set to the inner radius of the ring. On the inside, we assume that our icy dust particles sublimate when they reach the sublimation temperature of water ice, i.e. 100~K in vacuum. This effectively cuts off the inner disk at an inner radius of $\sim$35~AU.

\subsection{Fitting approach}

Model images in each of the five {\it Herschel} bands were created using the radiative-transfer code MCMax \citep{min09}. The code was modified to allow for an eccentric belt, with the star in one of the foci. The model images are convolved with the measured PSF of the PACS and SPIRE instruments at each observational wavelength. PACS and SPIRE images are obtained in broad-band filters. It is therefore not correct to directly compare the monochromatic model images with the observations. We take the filter transmission into account by computing model images at three wavelengths within each of the PACS and SPIRE bands\footnote{Three wavelengths per band is sufficient to sample the smooth wavelength-dependence of the model within the band, while keeping the computational time limited.}. The three images, weighted with the filter transmission curve, are then combined into a single image. The combined image represents the band-integrated model image.

The images of the three geometric components are stored separately. Hence, the images of the belt, the inner disk and the unresolved point source of a given model can be independently scaled to the observations.
We start with the PACS 70~\mic image and determine the best-fit linear combination of the three components using a non-negative least-squares (NNLS) fitting routine \citep{lawson74}. Because of the detected excess flux close to the star, the flux of the central source is not fixed to the stellar flux, but is allowed to exceed it. The circumstellar material around Fomalhaut is optically thin at all considered wavelengths. The scaling of the inner disk and belt is therefore equivalent to scaling the total dust mass in these components. The linear combination of the three geometric components hence provides the best possible fit to the 70~\mic image, for each model in the grid.

Two quantities are computed to assess the goodness-of-fit. First, a
Pearson chi-square, summed over all spatial pixels where the surface brightness is larger than 20$\sigma$, judges the general accordance between the linearly scaled model and the observation. Second, the maximum of the norm of the residuals is computed, to control the most extreme deviation between model and observation. Similar values were determined for the {\it Herschel} images at longer wavelengths. While these images are not used in the NNLS fitting routine, the agreement between model and observations is checked {\em a posteriori} and taken into account to determine the best-fit model in the grid.

\subsection{Best model}

\begin{figure}
\centering
\includegraphics[width=\columnwidth]{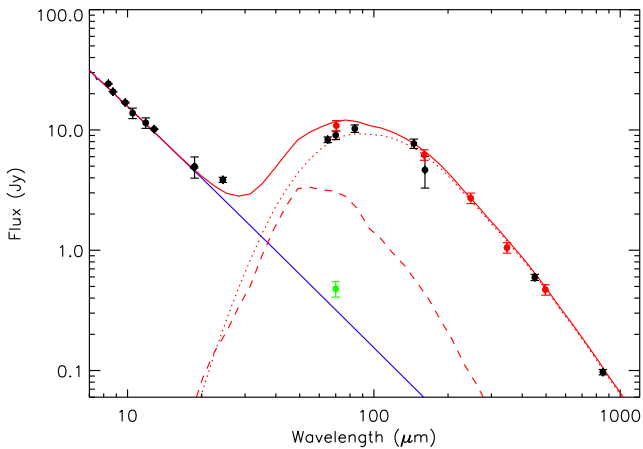}
\caption{The infrared SED of the Fomalhaut system. Black dots are photometric measurements from the literature, red dots the {\it Herschel} photometry. The full red line represents the best model, the dotted and dashed red lines indicate the contribution of the large grains and blow-out grains respectively. The blue line is the stellar photosphere model. The green dot indicates the total flux of the central point source in the 70~\mic image, and shows the excess of the stellar photosphere. The excess-to-stellar luminosity ratio, estimated from the model, is $7.4 \times 10^{-5}$, of which 27\% is produced by blow-out grains.}
           \label{fom_sed}%
\end{figure}

The best-fit model has a Pearson chi-square value, summed over all five images, of 0.42. The maximum pixel-to-pixel residual occurs in the 250~\mic image, where it amounts to 24\% of the maximum surface brightness of the observation. The model images and residuals are shown in Fig.~\ref{fom_images}. The general accordance between model and observations is very good. The most significant residual structure is seen in the 250~\mic image, where the belt appears to be more extended than the model predicts.

The best model has a debris source ring located between $r_1$ = 133~AU and $r_2$ = 153~AU. The belt contains a total mass of $8 \times 10^{25}$~g in dust grains up to sizes of 5000~\mic. The replenishment time, determining the amount of blow-out grains present in the system, is very short, approximately 1700~yr, leading to a total mass of $3 \times 10^{24}$ g in grains with sizes below 13~\mic. The inner disk contains $4 \times 10^{25}$~g of dust grains, and is responsible for 21$-$29\% of the flux in the {\it Herschel} images (Table~\ref{fom_phot}). Its surface density increases outward, linear with $r$, to the inner radius of the ring. The central unresolved excess-emission component has an integrated flux that amounts to 50\% of that of the star at 70~\mic (Table~\ref{fom_phot}). 

The grain properties that best fit the SED and images are icy grains with a vacuum fraction of 25\% by volume. However, we find little difference in total dust mass in blow-out grains -and hence in replenishment time- between models with different grain properties. This was to be expected; The key physical parameter that determines the thermal and dynamical properties of the dust particle is not the size, but the absorption efficiency. The grain model merely couples the absorption efficiency to a specific grain size, which is different for e.g. a different degree of porosity. Although grain models link different grain sizes to a similar absorption efficiency, it is this latter quantity that determines the appearance of the model. We conclude there is a significant amount of mass contained in grains with sizes smaller than the blow-out size in the Fomalhaut system. This conclusion is robust against assumptions for the grain structure and composition.

\section{Results}

\subsection{Hot excess emission}

\begin{figure}
\centering
\includegraphics[width=\columnwidth]{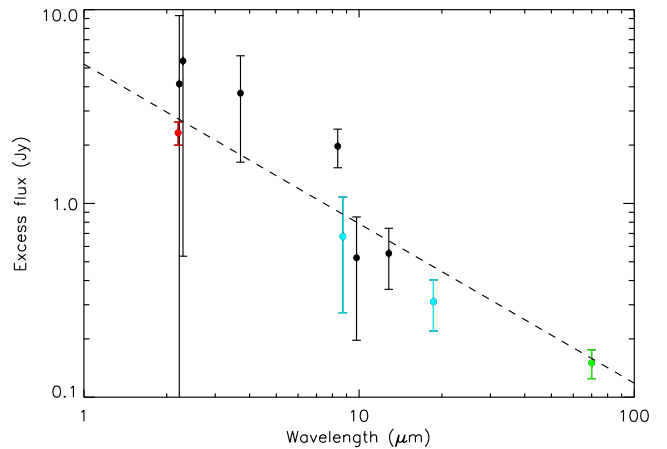}
\caption{The SED of the excess emission close to the star. Black dots are ESO photometry, blue dots AKARI photometry. The red dot refers to the near-IR interferometric
measurement \citep{absil09}, the green dot to the unresolved excess emission at the location of the star in the {\it Herschel} PACS 70\,\mic image. The dashed line represents $F_\nu \propto \nu^{0.8}$.}
           \label{fom_excess}%
\end{figure}

Excess emission close to the star has been reported at near-infrared wavelengths \citep{absil09}. The SED of our model shows that the emission from the inner disk and belt only becomes visible at wavelengths beyond 20~\mic. Closer inspection of the literature photometry shows that there is indeed a continuous excess above the stellar photospheric flux in the near- to mid-infrared wavelength range. In Fig.~\ref{fom_excess}, the SED of this excess is shown. The PACS 70~\mic photometric data point of the on-star excess is added. The error bars include the measurement uncertainties as listed in the different catalogues, but also the spread induced by uncertainties in the stellar model flux (typically 2\% at these wavelengths). We note that we have integrated the stellar spectrum within each of the photometric filters before computing the residual excess flux, i.e., the excess SED is not affected by a spurious color correction.

The spectral shape of the excess cannot be represented by a single-temperature blackbody. This could point to the presence of hot dust that is not confined to a narrow asteroid belt, but is distributed over a range of distances close to the star ($<$20~AU). Alternatively, the excess may also be due to emitting gas. The slope of the excess emission is $F_\nu \propto \nu^{0.8 \pm 0.1}$, close to the $\nu^{0.6}$ behavior expected for free-free emission from a spherical wind with constant outflow velocity around a hot star, as derived by \citet{panagia75}. Using their equation (24), a reasonable stellar mass-loss rate of $10^{-8}$~M$_\odot$\,yr$^{-1}$ is computed for Fomalhaut. Although \citet{absil09} rejected free-free emission as the origin of the near-infrared excess based on the absence of an emission component or variability in the H$\alpha$ line, our determination of the slope of the SED may revive the discussion on the origin of the excess. Further analysis is however beyond the scope of this paper.

\subsection{Cometary debris} 

The temperature of the particles in the belt, ranging from 40 K to 100 K
in our model, together with their now known location, requires the presence of particles with the absorption characteristics of small grains with sizes below the blow-out size. Paradoxically, HST imaging showed that the dust grains in the belt only scatter a very small fraction of the stellar light they receive into our line of sight \citep{kalas05}. This was interpreted as being due to the highly anisotropic scattering of compact particles larger than 100~\mic \citep{min10}. Very small compact particles, which have the required absorption properties to fit the temperature of the belt in our images, are incompatible with this very low scattering efficiency. Vice versa, the thermal properties of large compact 100~\mic particles are inconsistent with the observed thermal emission in the far-infrared.

To match both the scattered light images and our thermal infrared images, we need grains that absorb and emit like small grains, but scatter like large particles. We suggest here that this combination is characteristic of fluffy aggregates, similar to cometary interplanetary dust particles in our Solar System \citep{fraundorf82}. Such aggregates behave in many aspects as loosely bound collections of small monomers, and thus can have the absorption properties of small grains, while exhibiting the scattering anisotropy expected from large particles \citep{volten07}. The resulting highly anisotropic scattering ensures that most of the light is not scattered into our line of sight, giving the particles a very low {\em apparent} scattering cross section.

\citet{lebouquin09} have resolved the stellar rotation axis using near-infrared interferometry. Based on the orbital direction of Fomalhaut~b (i.e., counterclockwise), they argue that the eastern side of the belt is in fact the far side. This side is the bright side in scattered light, and hence the authors conclude that the particles are predominantly backscattering, an observation that was elaborated upon by \citet{min10}. However, two assumptions are crucial in the determination of the belt orientation. First, Fomalhaut~b must be a real planet in orbit. Second, the rotation axis of the planet's orbit and of the star must have the same direction, i.e., the planet must have a prograde orbit. If Fomalhaut~b is not a planet \citep[and there is reason for skepticism, see][]{janson12}, the far side of the belt cannot be determined from the available observations. If the planet is real, but has a retrograde orbit, the orientation is reversed, and the bright side would be the nearest side of the belt. In the latter case, the particles would be predominantly forward scattering.

Apart from these considerations on the orientation of the Fomalhaut system, our "fluffy aggregate" suggestion is consistent with both forward and backward scattering, or at least not in conflict with either one. Large aggregates have the scattering properties of large grains, with a strong diffraction spike outside of the observable range of scattering angles ($\theta \in [90^\circ - i = 25^\circ, 90^\circ + i = 155^\circ]$). Most of the stellar light would be scattered forward, but in a direction close to the plane of the belt. Whether the belt is dominated by backward or forward scattering at the observable scattering angles, would then depend on the specific scattering properties of the building blocks of the fluffy aggregate. Our suggestion does exclude large compact grains, because these grains do not have the thermal properties needed to explain the far-infrared {\it Herschel} images and the SED.

Fluffy aggregates have been invoked to explain the observed polarization properties of the debris disk around AU~Mic \citep{graham07}. Here we used a different method to come to a similar conclusion on the morphology of the dust grains in the Fomalhaut system.

\subsection{Mass loss in the belt}

The amount of blow-out particles needed to reproduce the images (Fig.~\ref{fom_images}) and the SED (Fig.~\ref{fom_sed}) can be directly converted into a mass-loss rate from the belt and, equivalently, to a dust production rate assuming a steady state. We find a mass-loss rate of $2 \times 10^{21}$ g yr$^{-1}$, corresponding to about two 10-km sized comets, or 2000 1-km-sized comets per day. 

We can further estimate the total number of comets in the ring required to generate such a high collision rate. Assuming that the ring ranges from $r_1$ = 133 AU to $r_2$ = 153 AU and has a normalized half-height $h = H / (2r)$, its volume is $V = \frac{4 \pi}{3} h (r_2^3 - r_1^3)$. The collision rate in the belt is then given by $N_\mathrm{c}^2 v_\mathrm{coll} \sigma_\mathrm{coll} / (2V)$, where $N_\mathrm{c}$ is the number of comets in the belt, $v_\mathrm{coll}$ is the relative velocity between comets, and $\sigma_\mathrm{coll} = 4 \pi r_\mathrm{c}^2$ is the collisional cross section of two comets with radius $r_\mathrm{c}$. Each collision destroys two comets with mass $\frac{4 \pi}{3} r_\mathrm{c}^3 \rho_\mathrm{c}$, where $\rho_\mathrm{c}$ is the specific density of a comet. The mass-loss rate is then \[ \frac{dM}{dt} = \frac{16 \pi^2 N_\mathrm{c}^2 v_\mathrm{coll} \rho_\mathrm{c} r_\mathrm{c}^5}{3V}. \]

We take $h = 0.1$, $v_\mathrm{coll} \approx h v_\mathrm{orbit}$ = 0.5 km/s \citep{dominik03c}, and $\rho_\mathrm{c}$ = 0.6 g cm$^{-3}$ \citep{britt06}, and find that a population of $2.6 \times 10^{11}$ 10-km-sized comets, or $8.3 \times 10^{13}$ 1-km-sized comets are required in the ring to produce the
observed collision rate. The Oort Cloud of the Solar System contains an estimated $10^{12} - 10^{13}$ comets which presumably came from the planetary system \citep{weissman91}, so these numbers are
not unreasonable. The total mass of the Fomalhaut belt is about 110 Earth masses, which
can be compared to the approximately 30 Earth masses required in the primordial Kuiper
Belt to allow for the formation of Pluto and other trans-Neptunian dwarf planets \citep{stern96}.

Interestingly, we find the same mass in Fomalhaut's belt if we extrapolate the size
distribution from the largest grains in our model (5000~\mic) to the cometary regime,
providing a second independent measurement of the same quantity. Crowding of
planetesimals into a narrow belt through resonance capture by an outward-migrating planet
can conceivably produce belts with high planetesimal content, and may have happened in
our own Solar System's Kuiper Belt during Neptune's migration \citep{hahn99}. Large amounts of
mass in the outer regions of disks around A-type stars like Fomalhaut might explain the
existence of planets found far away from the host star \citep{kalas08,marois08}, as in-situ planet formation could then have been possible.

\section{Conclusion}

The belt around Fomalhaut is a prototypical example of a steady-state
collisional cascade. The high mass in blow-out grains evidences the extreme activity of the
system. The dust particles are fluffy aggregates, similar in morphology to the cometary
interplanetary dust particles in our Solar System. This indicates that the planetesimals at the top of the collisional cascade are in fact comets.

\begin{acknowledgements}
BV, CW, JB, WDM, KE, RH, CJ and PR acknowledge funding from the Belgian Federal Science Policy Office via the ESA-PRODEX office.
\end{acknowledgements}

\bibliographystyle{aa}
\bibliography{/lhome/bram/REFERENCES/references.bib}

\begin{thebibliography}{49}
\expandafter\ifx\csname natexlab\endcsname\relax\def\natexlab#1{#1}\fi

\bibitem[{{Absil} {et~al.}(2009){Absil}, {Mennesson}, {Le Bouquin}, {Di Folco},
  {Kervella}, \& {Augereau}}]{absil09}
{Absil}, O., {Mennesson}, B., {Le Bouquin}, J.-B., {et~al.} 2009, \apj, 704,
  150

\bibitem[{{Aumann}(1985)}]{aumann85}
{Aumann}, H.~H. 1985, \pasp, 97, 885

\bibitem[{{Britt} {et~al.}(2006){Britt}, {Consolmagno}, \& {Merline}}]{britt06}
{Britt}, D.~T., {Consolmagno}, G.~J., \& {Merline}, W.~J. 2006, in Lunar and
  Planetary Institute Science Conference Abstracts, Vol.~37, 37th Annual Lunar
  and Planetary Science Conference, ed. {S.~Mackwell \& E.~Stansbery}, 2214

\bibitem[{{Bruggeman}(1935)}]{bruggeman35}
{Bruggeman}, D.~A.~G. 1935, Annalen der Physik, 416, 636

\bibitem[{{Bryden} {et~al.}(2006){Bryden}, {Beichman}, {Trilling}, {Rieke},
  {Holmes}, {Lawler}, {Stapelfeldt}, {Werner}, {Gautier}, {Blaylock}, {Gordon},
  {Stansberry}, \& {Su}}]{bryden06}
{Bryden}, G., {Beichman}, C.~A., {Trilling}, D.~E., {et~al.} 2006, \apj, 636,
  1098

\bibitem[{{Burns} {et~al.}(1979){Burns}, {Lamy}, \& {Soter}}]{burns79}
{Burns}, J.~A., {Lamy}, P.~L., \& {Soter}, S. 1979, \icarus, 40, 1

\bibitem[{{Castelli} \& {Kurucz}(2003)}]{castelli03}
{Castelli}, F. \& {Kurucz}, R.~L. 2003, in IAU Symposium, Vol. 210, Modelling
  of Stellar Atmospheres, ed. {N.~Piskunov, W.~W.~Weiss, \& D.~F.~Gray}, 20P

\bibitem[{{Chiang} {et~al.}(2009){Chiang}, {Kite}, {Kalas}, {Graham}, \&
  {Clampin}}]{chiang09}
{Chiang}, E., {Kite}, E., {Kalas}, P., {Graham}, J.~R., \& {Clampin}, M. 2009,
  \apj, 693, 734

\bibitem[{{Degroote} {et~al.}(2011){Degroote}, {Acke}, {Samadi}, {Aerts},
  {Kurtz}, {Noels}, {Miglio}, {Montalb{\'a}n}, {Bloemen}, {Baglin}, {Baudin},
  {Catala}, {Michel}, \& {Auvergne}}]{degroote11}
{Degroote}, P., {Acke}, B., {Samadi}, R., {et~al.} 2011, \aap, 536, A82

\bibitem[{{Di Folco} {et~al.}(2004){Di Folco}, {Th{\'e}venin}, {Kervella},
  {Domiciano de Souza}, {Coud{\'e} du Foresto}, {S{\'e}gransan}, \&
  {Morel}}]{difolco04}
{Di Folco}, E., {Th{\'e}venin}, F., {Kervella}, P., {et~al.} 2004, \aap, 426,
  601

\bibitem[{{Dohnanyi}(1969)}]{dohnanyi69}
{Dohnanyi}, J.~S. 1969, \jgr, 74, 2531

\bibitem[{{Dominik} \& {Decin}(2003)}]{dominik03c}
{Dominik}, C. \& {Decin}, G. 2003, \apj, 598, 626

\bibitem[{{Fraundorf} {et~al.}(1982){Fraundorf}, {Walker}, \&
  {Brownlee}}]{fraundorf82}
{Fraundorf}, P., {Walker}, R.~M., \& {Brownlee}, D.~E. 1982, in IAU Colloq. 61:
  Comet Discoveries, Statistics, and Observational Selection, ed.
  {L.~L.~Wilkening}, 383--409

\bibitem[{{Graham} {et~al.}(2007){Graham}, {Kalas}, \& {Matthews}}]{graham07}
{Graham}, J.~R., {Kalas}, P.~G., \& {Matthews}, B.~C. 2007, \apj, 654, 595

\bibitem[{{Grevesse} \& {Sauval}(1998)}]{grevesse98}
{Grevesse}, N. \& {Sauval}, A.~J. 1998, \ssr, 85, 161

\bibitem[{{Griffin} {et~al.}(2010){Griffin}, {Abergel}, {Abreu}, {Ade},
  {Andr{\'e}}, {Augueres}, {Babbedge}, {Bae}, {Baillie}, {Baluteau}, {Barlow},
  {Bendo}, {Benielli}, {Bock}, {Bonhomme}, {Brisbin}, {Brockley-Blatt},
  {Caldwell}, {Cara}, {Castro-Rodriguez}, {Cerulli}, {Chanial}, {Chen},
  {Clark}, {Clements}, {Clerc}, {Coker}, {Communal}, {Conversi}, {Cox},
  {Crumb}, {Cunningham}, {Daly}, {Davis}, {de Antoni}, {Delderfield}, {Devin},
  {di Giorgio}, {Didschuns}, {Dohlen}, {Donati}, {Dowell}, {Dowell}, {Duband},
  {Dumaye}, {Emery}, {Ferlet}, {Ferrand}, {Fontignie}, {Fox}, {Franceschini},
  {Frerking}, {Fulton}, {Garcia}, {Gastaud}, {Gear}, {Glenn}, {Goizel},
  {Griffin}, {Grundy}, {Guest}, {Guillemet}, {Hargrave}, {Harwit}, {Hastings},
  {Hatziminaoglou}, {Herman}, {Hinde}, {Hristov}, {Huang}, {Imhof}, {Isaak},
  {Israelsson}, {Ivison}, {Jennings}, {Kiernan}, {King}, {Lange}, {Latter},
  {Laurent}, {Laurent}, {Leeks}, {Lellouch}, {Levenson}, {Li}, {Li},
  {Lilienthal}, {Lim}, {Liu}, {Lu}, {Madden}, {Mainetti}, {Marliani}, {McKay},
  {Mercier}, {Molinari}, {Morris}, {Moseley}, {Mulder}, {Mur}, {Naylor},
  {Nguyen}, {O'Halloran}, {Oliver}, {Olofsson}, {Olofsson}, {Orfei}, {Page},
  {Pain}, {Panuzzo}, {Papageorgiou}, {Parks}, {Parr-Burman}, {Pearce},
  {Pearson}, {P{\'e}rez-Fournon}, {Pinsard}, {Pisano}, {Podosek}, {Pohlen},
  {Polehampton}, {Pouliquen}, {Rigopoulou}, {Rizzo}, {Roseboom}, {Roussel},
  {Rowan-Robinson}, {Rownd}, {Saraceno}, {Sauvage}, {Savage}, {Savini},
  {Sawyer}, {Scharmberg}, {Schmitt}, {Schneider}, {Schulz}, {Schwartz},
  {Shafer}, {Shupe}, {Sibthorpe}, {Sidher}, {Smith}, {Smith}, {Smith},
  {Spencer}, {Stobie}, {Sudiwala}, {Sukhatme}, {Surace}, {Stevens}, {Swinyard},
  {Trichas}, {Tourette}, {Triou}, {Tseng}, {Tucker}, {Turner}, {Vaccari},
  {Valtchanov}, {Vigroux}, {Virique}, {Voellmer}, {Walker}, {Ward}, {Waskett},
  {Weilert}, {Wesson}, {White}, {Whitehouse}, {Wilson}, {Winter}, {Woodcraft},
  {Wright}, {Xu}, {Zavagno}, {Zemcov}, {Zhang}, \& {Zonca}}]{griffin10}
{Griffin}, M.~J., {Abergel}, A., {Abreu}, A., {et~al.} 2010, \aap, 518, L3

\bibitem[{{Grigorieva} {et~al.}(2007){Grigorieva}, {Th{\'e}bault},
  {Artymowicz}, \& {Brandeker}}]{grigorieva07}
{Grigorieva}, A., {Th{\'e}bault}, P., {Artymowicz}, P., \& {Brandeker}, A.
  2007, \aap, 475, 755

\bibitem[{{Hahn} \& {Malhotra}(1999)}]{hahn99}
{Hahn}, J.~M. \& {Malhotra}, R. 1999, \aj, 117, 3041

\bibitem[{{Holland} {et~al.}(2003){Holland}, {Greaves}, {Dent}, {Wyatt},
  {Zuckerman}, {Webb}, {McCarthy}, {Coulson}, {Robson}, \& {Gear}}]{holland03}
{Holland}, W.~S., {Greaves}, J.~S., {Dent}, W.~R.~F., {et~al.} 2003, \apj, 582,
  1141

\bibitem[{{Ishihara} {et~al.}(2010){Ishihara}, {Onaka}, {Kataza}, {Salama},
  {Alfageme}, {Cassatella}, {Cox}, {Garc{\'{\i}}a-Lario}, {Stephenson},
  {Cohen}, {Fujishiro}, {Fujiwara}, {Hasegawa}, {Ita}, {Kim}, {Matsuhara},
  {Murakami}, {M{\"u}ller}, {Nakagawa}, {Ohyama}, {Oyabu}, {Pyo}, {Sakon},
  {Shibai}, {Takita}, {Tanab{\'e}}, {Uemizu}, {Ueno}, {Usui}, {Wada},
  {Watarai}, {Yamamura}, \& {Yamauchi}}]{ishihara10}
{Ishihara}, D., {Onaka}, T., {Kataza}, H., {et~al.} 2010, \aap, 514, A1

\bibitem[{{Janson} {et~al.}(2012){Janson}, {Carson}, {Lafreniere}, {Spiegel},
  {Bent}, \& {Wong}}]{janson12}
{Janson}, M., {Carson}, J., {Lafreniere}, D., {et~al.} 2012, ArXiv:1201.4388

\bibitem[{{Kalas} {et~al.}(2008){Kalas}, {Graham}, {Chiang}, {Fitzgerald},
  {Clampin}, {Kite}, {Stapelfeldt}, {Marois}, \& {Krist}}]{kalas08}
{Kalas}, P., {Graham}, J.~R., {Chiang}, E., {et~al.} 2008, Science, 322, 1345

\bibitem[{{Kalas} {et~al.}(2005){Kalas}, {Graham}, \& {Clampin}}]{kalas05}
{Kalas}, P., {Graham}, J.~R., \& {Clampin}, M. 2005, \nat, 435, 1067

\bibitem[{{Kalas} {et~al.}(2011){Kalas}, {Graham}, {Fitzgerald}, \&
  {Clampin}}]{kalas11}
{Kalas}, P., {Graham}, J.~R., {Fitzgerald}, M.~P., \& {Clampin}, M. 2011, in
  Bulletin of the American Astronomical Society, Vol.~43, American Astronomical
  Society Meeting Abstracts \#217, 302.02

\bibitem[{{Krivov}(2010)}]{krivov10}
{Krivov}, A.~V. 2010, Research in Astronomy and Astrophysics, 10, 383

\bibitem[{{Lawson} \& {Hanson}(1974)}]{lawson74}
{Lawson}, C.~L. \& {Hanson}, R.~J. 1974, {Solving least squares problems}, ed.
  {Lawson, C.~L.~\& Hanson, R.~J.}

\bibitem[{{Le Bouquin} {et~al.}(2009){Le Bouquin}, {Absil}, {Benisty}, {Massi},
  {M{\'e}rand}, \& {Stefl}}]{lebouquin09}
{Le Bouquin}, J.-B., {Absil}, O., {Benisty}, M., {et~al.} 2009, \aap, 498, L41

\bibitem[{{Marois} {et~al.}(2008){Marois}, {Macintosh}, {Barman}, {Zuckerman},
  {Song}, {Patience}, {Lafreni{\`e}re}, \& {Doyon}}]{marois08}
{Marois}, C., {Macintosh}, B., {Barman}, T., {et~al.} 2008, Science, 322, 1348

\bibitem[{{Min} {et~al.}(2009){Min}, {Dullemond}, {Dominik}, {de Koter}, \&
  {Hovenier}}]{min09}
{Min}, M., {Dullemond}, C.~P., {Dominik}, C., {de Koter}, A., \& {Hovenier},
  J.~W. 2009, \aap, 497, 155

\bibitem[{{Min} {et~al.}(2011){Min}, {Dullemond}, {Kama}, \& {Dominik}}]{min11}
{Min}, M., {Dullemond}, C.~P., {Kama}, M., \& {Dominik}, C. 2011, \icarus, 212,
  416

\bibitem[{{Min} {et~al.}(2005){Min}, {Hovenier}, \& {de Koter}}]{min05}
{Min}, M., {Hovenier}, J.~W., \& {de Koter}, A. 2005, \aap, 432, 909

\bibitem[{{Min} {et~al.}(2010){Min}, {Kama}, {Dominik}, \& {Waters}}]{min10}
{Min}, M., {Kama}, M., {Dominik}, C., \& {Waters}, L.~B.~F.~M. 2010, \aap, 509,
  L6

\bibitem[{{Ott}(2010)}]{ott10}
{Ott}, S. 2010, in Astronomical Society of the Pacific Conference Series, Vol.
  434, Astronomical Data Analysis Software and Systems XIX, ed. {Y.~Mizumoto,
  K.-I.~Morita, \& M.~Ohishi}, 139

\bibitem[{{Panagia} \& {Felli}(1975)}]{panagia75}
{Panagia}, N. \& {Felli}, M. 1975, \aap, 39, 1

\bibitem[{{Pilbratt} {et~al.}(2010){Pilbratt}, {Riedinger}, {Passvogel},
  {Crone}, {Doyle}, {Gageur}, {Heras}, {Jewell}, {Metcalfe}, {Ott}, \&
  {Schmidt}}]{pilbratt10}
{Pilbratt}, G.~L., {Riedinger}, J.~R., {Passvogel}, T., {et~al.} 2010, \aap,
  518, L1

\bibitem[{{Poglitsch} {et~al.}(2010){Poglitsch}, {Waelkens}, {Geis},
  {Feuchtgruber}, {Vandenbussche}, {Rodriguez}, {Krause}, {Renotte}, {van
  Hoof}, {Saraceno}, {Cepa}, {Kerschbaum}, {Agn{\`e}se}, {Ali}, {Altieri},
  {Andreani}, {Augueres}, {Balog}, {Barl}, {Bauer}, {Belbachir}, {Benedettini},
  {Billot}, {Boulade}, {Bischof}, {Blommaert}, {Callut}, {Cara}, {Cerulli},
  {Cesarsky}, {Contursi}, {Creten}, {De Meester}, {Doublier}, {Doumayrou},
  {Duband}, {Exter}, {Genzel}, {Gillis}, {Gr{\"o}zinger}, {Henning},
  {Herreros}, {Huygen}, {Inguscio}, {Jakob}, {Jamar}, {Jean}, {de Jong},
  {Katterloher}, {Kiss}, {Klaas}, {Lemke}, {Lutz}, {Madden}, {Marquet},
  {Martignac}, {Mazy}, {Merken}, {Montfort}, {Morbidelli}, {M{\"u}ller},
  {Nielbock}, {Okumura}, {Orfei}, {Ottensamer}, {Pezzuto}, {Popesso},
  {Putzeys}, {Regibo}, {Reveret}, {Royer}, {Sauvage}, {Schreiber}, {Stegmaier},
  {Schmitt}, {Schubert}, {Sturm}, {Thiel}, {Tofani}, {Vavrek}, {Wetzstein},
  {Wieprecht}, \& {Wiezorrek}}]{poglitsch10}
{Poglitsch}, A., {Waelkens}, C., {Geis}, N., {et~al.} 2010, \aap, 518, L2

\bibitem[{{Rufener}(1988)}]{rufener88}
{Rufener}, F. 1988, {Catalogue of stars measured in the Geneva Observatory
  photometric system : 4 : 1988}

\bibitem[{{Stapelfeldt} {et~al.}(2004){Stapelfeldt}, {Holmes}, {Chen}, {Rieke},
  {Su}, {Hines}, {Werner}, {Beichman}, {Jura}, {Padgett}, {Stansberry},
  {Bendo}, {Cadien}, {Marengo}, {Thompson}, {Velusamy}, {Backus}, {Blaylock},
  {Egami}, {Engelbracht}, {Frayer}, {Gordon}, {Keene}, {Latter}, {Megeath},
  {Misselt}, {Morrison}, {Muzerolle}, {Noriega-Crespo}, {Van Cleve}, \&
  {Young}}]{stapelfeldt04}
{Stapelfeldt}, K.~R., {Holmes}, E.~K., {Chen}, C., {et~al.} 2004, \apjs, 154,
  458

\bibitem[{{Stern}(1996)}]{stern96}
{Stern}, S.~A. 1996, \aj, 112, 1203

\bibitem[{{Su} {et~al.}(2006){Su}, {Rieke}, {Stansberry}, {Bryden},
  {Stapelfeldt}, {Trilling}, {Muzerolle}, {Beichman}, {Moro-Martin}, {Hines},
  \& {Werner}}]{su06}
{Su}, K.~Y.~L., {Rieke}, G.~H., {Stansberry}, J.~A., {et~al.} 2006, \apj, 653,
  675

\bibitem[{{Swinyard} {et~al.}(2010){Swinyard}, {Ade}, {Baluteau}, {Aussel},
  {Barlow}, {Bendo}, {Benielli}, {Bock}, {Brisbin}, {Conley}, {Conversi},
  {Dowell}, {Dowell}, {Ferlet}, {Fulton}, {Glenn}, {Glauser}, {Griffin},
  {Griffin}, {Guest}, {Imhof}, {Isaak}, {Jones}, {King}, {Leeks}, {Levenson},
  {Lim}, {Lu}, {Makiwa}, {Naylor}, {Nguyen}, {Oliver}, {Panuzzo},
  {Papageorgiou}, {Pearson}, {Pohlen}, {Polehampton}, {Pouliquen},
  {Rigopoulou}, {Ronayette}, {Roussel}, {Rykala}, {Savini}, {Schulz},
  {Schwartz}, {Shupe}, {Sibthorpe}, {Sidher}, {Smith}, {Spencer}, {Trichas},
  {Triou}, {Valtchanov}, {Wesson}, {Woodcraft}, {Xu}, {Zemcov}, \&
  {Zhang}}]{swinyard10}
{Swinyard}, B.~M., {Ade}, P., {Baluteau}, J.-P., {et~al.} 2010, \aap, 518, L4

\bibitem[{{Telesco} {et~al.}(2000){Telesco}, {Fisher}, {Pi{\~n}a}, {Knacke},
  {Dermott}, {Wyatt}, {Grogan}, {Holmes}, {Ghez}, {Prato}, {Hartmann}, \&
  {Jayawardhana}}]{telesco00}
{Telesco}, C.~M., {Fisher}, R.~S., {Pi{\~n}a}, R.~K., {et~al.} 2000, \apj, 530,
  329

\bibitem[{{van der Bliek} {et~al.}(1996){van der Bliek}, {Manfroid}, \&
  {Bouchet}}]{vanderbliek96}
{van der Bliek}, N.~S., {Manfroid}, J., \& {Bouchet}, P. 1996, \aaps, 119, 547

\bibitem[{{Volten} {et~al.}(2007){Volten}, {Mu{\~n}oz}, {Hovenier},
  {Rietmeijer}, {Nuth}, {Waters}, \& {van der Zande}}]{volten07}
{Volten}, H., {Mu{\~n}oz}, O., {Hovenier}, J.~W., {et~al.} 2007, \aap, 470, 377

\bibitem[{{Voshchinnikov} {et~al.}(2007){Voshchinnikov}, {Videen}, \&
  {Henning}}]{voshchinnikov07}
{Voshchinnikov}, N.~V., {Videen}, G., \& {Henning}, T. 2007, \ao, 46, 4065

\bibitem[{{Weissman}(1991)}]{weissman91}
{Weissman}, P.~R. 1991, in Astrophysics and Space Science Library, Vol. 167,
  IAU Colloq. 116: Comets in the post-Halley era, ed. {R.~L.~Newburn Jr.,
  M.~Neugebauer, \& J.~Rahe}, 463--486

\bibitem[{{Wyatt}(2008)}]{wyatt08}
{Wyatt}, M.~C. 2008, \araa, 46, 339

\bibitem[{{Wyatt} \& {Dent}(2002)}]{wyatt02}
{Wyatt}, M.~C. \& {Dent}, W.~R.~F. 2002, \mnras, 334, 589

\bibitem[{{Yamamura} {et~al.}(2010){Yamamura}, {Makiuti}, {Ikeda}, {Fukuda},
  {Oyabu}, {Koga}, \& {White}}]{yamamura10}
{Yamamura}, I., {Makiuti}, S., {Ikeda}, N., {et~al.} 2010, VizieR Online Data
  Catalog, 2298

\end{thebibliography}

\begin{appendix}
\section{Particle size and surface density distribution in the inner disk.\label{appA}}

Here we compute the size distribution of dust grains in the inner disk under the assumption that these grains are moving inward due to Poynting-Robertson (PR) drag, and that no collisions occur in the PR stream inward of the source region. Because the absolute scaling (i.e. total dust mass) is determined by the fitting routine (see Sect.~\ref{modeling}), we only consider the proportionalities of the different processes.

We assume that the ring itself is in collisional equilibrium, so that the lifetime of particles is dominated by collisions for all bound particles. Even then, there will be a narrow slice close to the inner boundary of the ring where the time scale for a particle to move out of the ring by PR drag is shorter than its collisional destruction time scale. We assume that the particle flux seeping into the inner disk is proportional to the number of particles of a given size in that slice. Note that the width of this slice depends on grain size.

In the collisional ring itself, we assume that the size distribution is given by $f(s) \propto s^\alpha$, where $\alpha = -3.5$ in case of collisional equilibrium. First, we estimate the collisional time scale for particles of size $s$. The most important collision partners are grains with sizes near $s$, in a size range $ds \propto s$ \citep{dominik03c}. If $v_\mathrm{coll}$ is the collision velocity and $\sigma_\mathrm{coll}$ is the collisional cross section, we see that the collisional time scales as $t_\mathrm{coll} \propto \left( f(s)\, s\, \sigma_\mathrm{coll} \right)^{-1}$. We assume that $v_\mathrm{coll}$ is independent of grain size, i.e. particles with different sizes have similar orbits. Because $\sigma_\mathrm{coll} \propto s^2$, we find that $t_\mathrm{coll} \propto s^{-(3+\alpha)}$.

PR drag pulls particles out of the ring. The inward velocity due to PR drag at a distance $r$
from the star is $v_\mathrm{PR} \propto \frac{r}{t_\mathrm{PR}} \propto \frac{r\, \beta(s)}{r^2} \propto \frac{\beta(s)}{r}$, where $\beta(s)$ is again the ratio of the radiation and
gravitational forces on a grain of size $s$. At the inner radius $r_1$ of the ring, $v_\mathrm{PR}(r_1) \propto \beta(s)$. We now assume that the particle flux through $r_1$ into the inner disk is proportional to the material available in the narrow slice $dr = v_\mathrm{PR}(r_1)\, t_\mathrm{coll}$ where the PR time scale is shorter than the collisional time scale. The particle flux is then proportional to $F(s) \propto f(s) dr = f(s)\, v_\mathrm{PR}(r_1)\, t_\mathrm{coll} \propto s^\alpha\, \beta(s)\, s^{-(3+\alpha)} \propto \beta(s) s^{-3}$.

Mass conservation in the flow through the inner disk then gives the surface density $\Sigma(s) = \frac{F(s)}{2 \pi r v_\mathrm{PR}}$, where we now have to consider that $v_\mathrm{PR}$ is no longer evaluated at the inner radius of the ring, but changes proportional to $\frac{\beta(s)}{r}$. We find that $\Sigma(s) \propto \frac{\beta(s) s^{-3}}{\beta(s)} = s^{-3}$. For an undisturbed PR flow, the surface density depends on the grain size, but not on the distance to the star. Based on this reasoning, we use a size distribution $f(s) \propto s^{-3}$ in the inner disk. Note, however, that in reality, the size distribution may be modified by interactions with a planetary system. Also other sources of dust, such as evaporating comets or an unresolved inner belt, would give rise to different size distributions.

\end{appendix}

\end{document}